\input{aipcheck}

\documentclass[
    ,final            % use final for the camera ready runs
  ]
  {aipproc}

\usepackage{amssymb}

\layoutstyle{8x11single}

\def\be{\begin{equation}}
\def\ee{\end{equation}}
\def\ba{\begin{eqnarray}}
\def\ea{\end{eqnarray}}

\def\sfrac#1#2{{\textstyle \frac{#1}{#2}}}

\begin{document}

\title{
$N^\ast(1535)$ electroproduction at high $Q^2$}

%12.39.Ki 	Relativistic quark model
%13.40.Gp 	Electromagnetic form factors
%14.20.Gk 	Baryon resonances (S=C=B=0)

\classification{12.39.Ki, 13.40.Gp, 14.20.Gk}
\keywords      {Covariant quark model, $S_{11}(1535)$ electroproduction,
Helicity amplitudes}

\author{G. Ramalho}{
  address={CFTP, IST, Universidade T\'ecnica de Lisboa,
Av.\ Rovisco Pais, 1049-001 Lisboa, Portugal}
}

\author{M.T. Pe\~na}{
   address={
CFTP, IST, Universidade T\'ecnica de Lisboa,
Av.\ Rovisco Pais, 1049-001 Lisboa, Portugal},
  altaddress={Physics Dept., IST, Universidade T\'ecnica de Lisboa,
Av.\ Rovisco Pais, 1049-001 Lisboa, Portugal},
}

\author{K. Tsushima}{
  address={
CSSM, School of Chemistry and Physics,
University of Adelaide, Adelaide SA 5005, Australia}
}

\begin{abstract}
A covariant spectator quark model is applied to study the
$\gamma N \to N^\ast(1535)$ reaction in the large $Q^2$ region.
Starting from the relation between the nucleon
and $N^\ast(1535)$ systems, the $N^\ast(1535)$
valence quark wave function
is determined
without the addition of any parameters.
The model is then used to calculate
the $\gamma N \to N^\ast(1535)$ transition form factors.
A very interesting, useful relation between the
$A_{1/2}$ and $S_{1/2}$ helicity amplitudes for $Q^2> 2$ GeV$^2$,
is also derived.
\end{abstract}

\maketitle

%%%%%%%%%%%%%%%%%%%%%%%%%%%%%%%%%%%%%%%%%%%%
%% MAINMATTER
%%%%%%%%%%%%%%%%%%%%%%%%%%%%%%%%%%%%%%%%%%%%

\section{Introduction}

Study of the meson-nucleon reactions is one of the
most important research topics associated with modern
accelerators like CEBAF at Jefferson Lab,
and defines new challenges for theoretical models.
Although the electroproduction of nucleon
resonances ($\gamma N \to N^\ast$) is expected
to be governed by the interaction of quarks and gluons (QCD)   
at very large momentum transfer squared $Q^2$,
at present one has to rely on some effective 
and phenomenological approaches such as effective
meson-baryon models~\cite{Burkert04}
at low $Q^2$, and/or constituent
quark models~\cite{S11,S11scaling,Capstick95}
at moderate and large $Q^2$, where
meson cloud effects are attenuated.

The $N^\ast(1535)$ resonance, identified as an $S_{11}$ state,
is particularly interesting among the many
experimentally observed nucleon resonances.
It is the chiral partner ($J^P=\sfrac{1}{2}^-$) of the nucleon,
and has strong decay channels for both $\pi N$ and $\eta N$.
It has been suggested that $N^\ast(1535)$
can be dynamically generated as a $K \Sigma$
quasi-bound state~\cite{Kaiser95,Jido08,EBAC}.
But it was also argued that pure valence quark effects
are important to explain its electromagnetic structure~\cite{Jido08}.
Furthermore, the $N^\ast(1535)$ resonance is also interesting
due to the closeness in its mass with that of the other
$S_{11}$ state, $N^\ast(1650)$, where both states can be regarded
as combinations of the quark core states of spin 1/2 and 3/2,
and the mass splitting is due to the color hyperfine interactions
between the quarks~\cite{Capstick95}.

To describe the $\gamma N \to N^\ast(1535)$ transition,
we use a covariant spectator quark model~\cite{Nucleon,Omega,Octet,NDelta}.
The model has been successfully applied
for studying the properties of nucleon~\cite{Nucleon,Octet,ExclusiveR},
$\Delta$~\cite{NDelta,NDeltaD,LatticeD,DeltaFF},
higher resonances~\cite{S11,Roper,Delta1600},
and also the electromagnetic transitions
in the lattice QCD regime \cite{Omega,LatticeD,Lattice}.
In the covariant spectator quark model a baryon is described
as a three-valence quark system with an on-shell quark-pair (diquark)
with mass $m_D$, while the remaining quark is off-shell and
free to interact with the electromagnetic fields.
The quark-diquark vertex is represented by a baryon $B$ wave function
$\Psi_B$ that effectively describes quark confinement~\cite{Nucleon}.
To represent the nucleon system,
we adopt the simplest structure given by a symmetric
and anti-symmetric combination of the diquark states,
combined to a relative S-state with
the remaining quark~\cite{Nucleon}:
\be
\Psi_N(P,k)= \frac{1}{\sqrt{2}}
\left[ \Phi_I^0 \Phi_S^0 + \Phi_I^1 \Phi_S^1
\right] \psi_N(P,k), 
\label{eqPsiN}
\ee
where $\Phi_{S}^{0,1}$ [$\Phi_{I}^{0,1}$] is the
spin [isospin] state which
corresponds to the diquark with the quantum number 0 or 1.
The function $\psi_N$ is a scalar wave function
which depends exclusively on $(P-k)^2$,
where $P$ ($k$) is the baryon (diquark) momentum.
The $N^\ast(1535)$ state has the same
isospin state but different spin state with that of the nucleon.
The $N^\ast(1535)$ spin state can have P-states in the relative quark-diquark
configuration and/or in the diquark system.
Assuming that the core spin 1/2 state is dominant~\cite{Capstick95}
and a point-like diquark (no internal P-states), one can
represent the $N^\ast(1535)$ wave function as
\be
\Psi_{S11} (P,k) = \frac{1}{\sqrt{2}}
\gamma_5
\left[ \Phi_I^0 X_\rho - \Phi_I^1 X_\lambda
\right] \psi_{S11}(P,k),
\label{eqPsiS11}
\ee
where $X_\rho$ and $X_\lambda$ are respectively
the anti-symmetric and symmetric spin states
with respect to the exchange of the quarks 1 and 2,
and $\psi_{S11}$ is the $N^\ast(1535)$ radial
wave function~\cite{S11}.

The scalar wave functions $\psi_B (B=N,S_{11})$ are represented
using the dimensionless variable $\chi_{_B}$:
\ba
\chi_{_B}= \frac{(M_B-m_D)^2-(P-k)^2}{M_Bm_D},
\hspace{1.cm}
\psi_B(P,k)&=&
\frac{N_B}{m_D(\beta_1 + \chi_{_B})(\beta_2+\chi_{_B})}.
\label{eqPsiB}
\ea
Note that, $\chi_{_B}$ contains a dependence on the baryon mass
$M_B$ that can be $M$ for the nucleon and $M_S$ for the $N^\ast(1535)$.
In the above equation, $N_B$ are the normalization constants,
and $\beta_{1}$ and $\beta_2$ are momentum range parameters
that regulate the short and long range behavior in
position space.
As we represent both the wave functions with the same
parameters for $N$ and $N^\ast(1535)$,
they have the same form in the respective rest frames,
apart from the orbital angular momenta.
This means that no extra parameters are required to
represent the $N^\ast(1535)$ state.
The analytic form~(\ref{eqPsiB}),
was chosen to reproduce the asymptotic form predicted
by pQCD for the nucleon form factors ($G_E,G_M \sim 1/Q^4$),
but also assures the expected pQCD behavior for the
$\gamma N \to N^\ast(1535)$ transition form factors~\cite{S11}.

The constituent quark electromagnetic current in the model is described by
\be
j_I^\mu =
\left( \sfrac{1}{6}f_{1+} + \sfrac{1}{2}f_{1-} \tau_3 \right)
\left(\gamma^\mu - \frac{\not\!q q^\mu}{q^2}\right)+
  \left( \sfrac{1}{6}f_{2+} + \sfrac{1}{2}f_{2-} \tau_3 \right)
 \frac{i \sigma^{\mu \nu}q_\nu}{2M},
\ee
where $\tau_3$ is the isospin projection operator.
To parameterize the electromagnetic structure
of the constituent quark in terms of the
quark form factors $f_{1\pm}$ and $f_{2\pm}$,
we adopt a vector meson dominance-based parametrization~\cite{Nucleon,Omega}.

The $\gamma N \to S_{11}$ transition in the model
is described by a relativistic impulse approximation
in terms of the initial $P_-$ and final $P_+$ baryon momenta 
with the diquark (spectator) on-mass-shell~\cite{S11,S11scaling}:
\ba
J^\mu &=&  3 \sum_{\Lambda} \int_k \bar \Psi_R (P_-,k)j_I^\mu \Psi_N(P_-,k),
\nonumber \\
&=& \bar u_S (P_+) \left[
\left(\gamma^\mu -\frac{\not \! q q^\mu}{q^2} \right)F_1^\ast(Q^2)
+ \frac{i\sigma^{\mu \nu} q_\nu}{M_S + M} F_2^\ast (Q^2)
\right]\gamma_5
u(P_-).
\ea
In the first line
%In the first equation
the sum is over the diquark states $\Lambda=\{s,\lambda\}$,
where $s$ and $\lambda=0,\pm1$ stand for
the scalar diquark
and the vector diquark polarizations, respectively,
and $\int_k$ is a covariant integral in the diquark momentum.
The factor 3 is due to the flavor symmetry.
In the second line the transition
form factors $F_1^\ast$ and  $F_2^\ast$ are defined
independent of the frame
using the Dirac spinors of the $S_{11}$ state ($u_S$) and nucleon ($u$).
The spin projection indices
are suppressed for simplicity.

%%%%%%%%%%%%%%%%%%%%%%%%%%%%%%%%%%%%%%%%%%%%
%% Sample figure:
%%
%% The option [height=...] scales the picture to the given height,
%% without it it would be printed at its nominal size
%%%%%%%%%%%%%%%%%%%%%%%%%%%%%%%%%%%%%%%%%%%%

\begin{figure}
% original size: [width=3.2in]
  \includegraphics[width=3.0in]{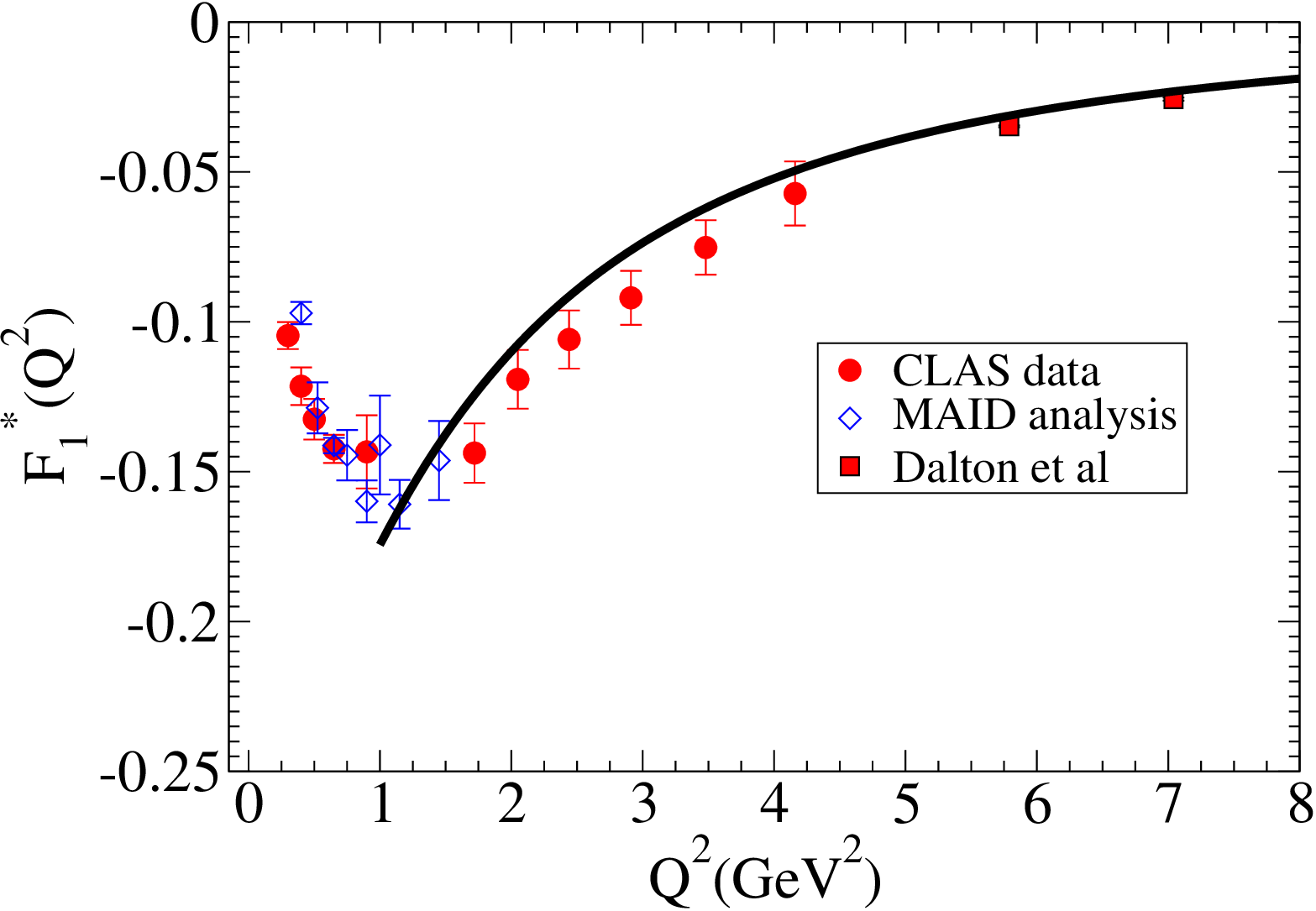}  \hspace{.1cm}
  \includegraphics[width=3.0in]{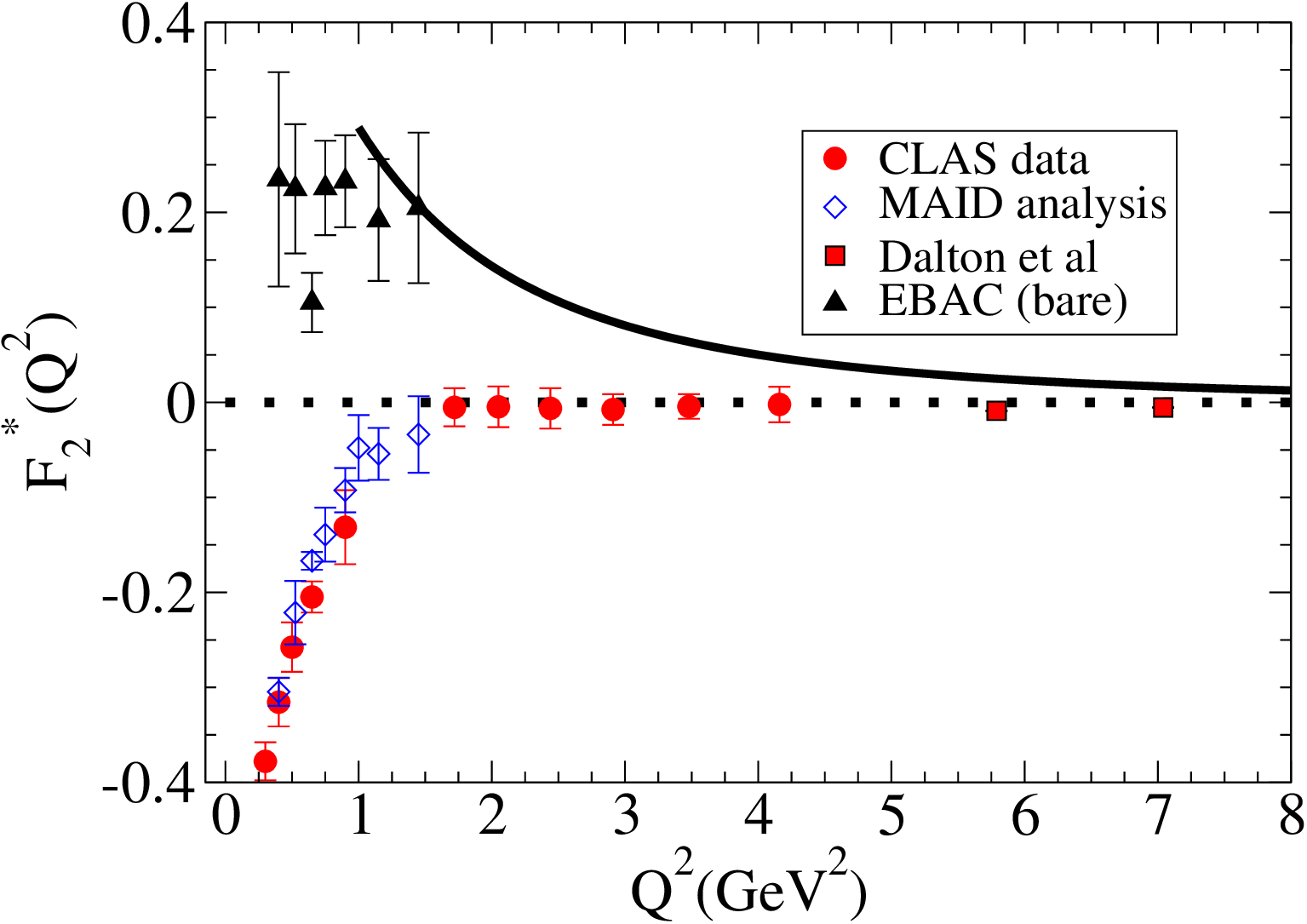}
  \caption{$\gamma N \to S_{11}(1525)$ transition form factors~\cite{S11}.
Left panel: model for $F_1^\ast$ compared with the data.
Right panel: model for $F_2^\ast$ compared with the data
and an estimate for the quark core contributions from EBAC~\cite{EBAC}.
Data are from Refs.~\cite{CLAS,MAID,Dalton09}.}
\label{figFF}
\end{figure}

\begin{figure}
% original size: [width=3.2in]
  \includegraphics[width=3.0in]{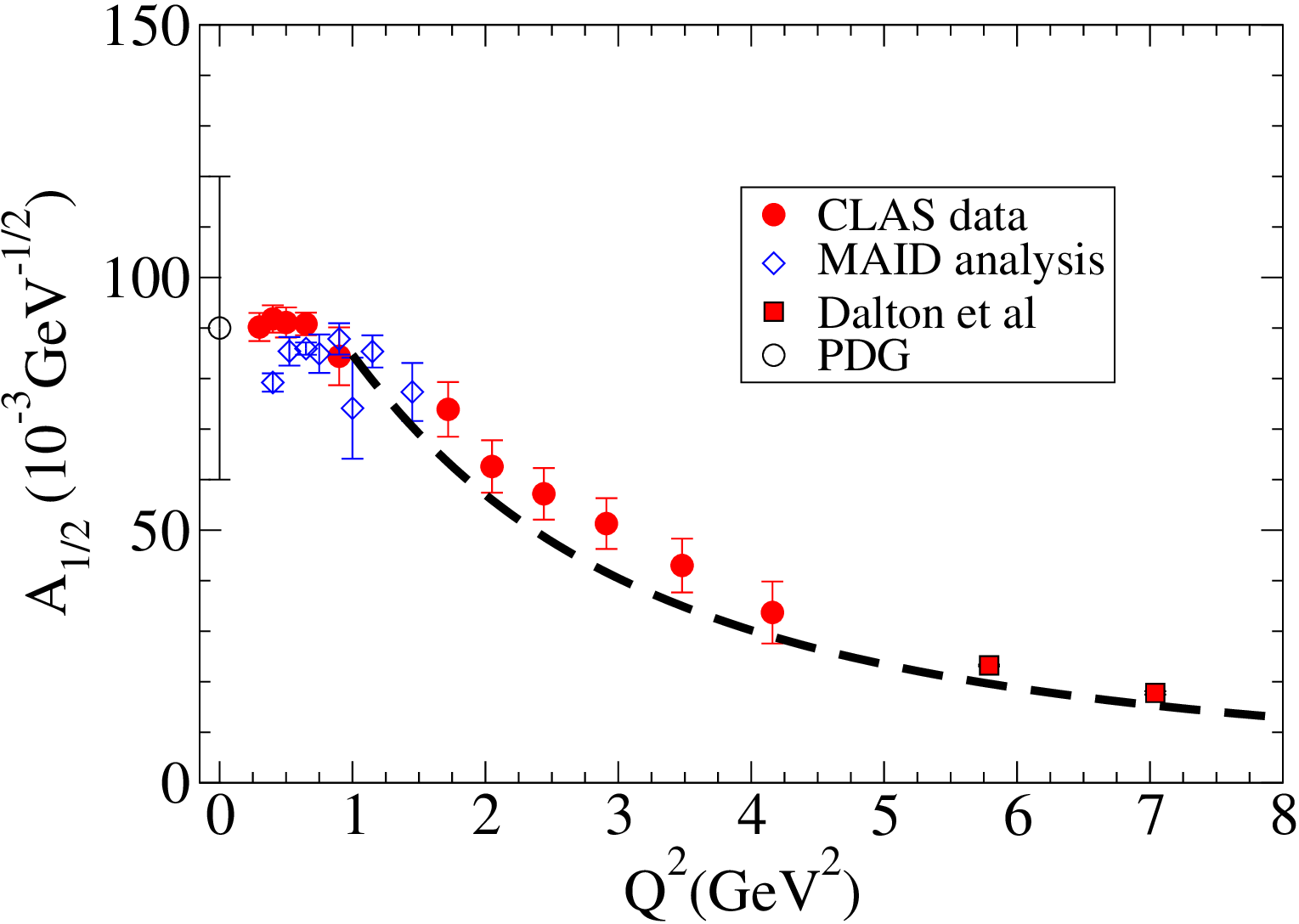}  \hspace{.1cm}
  \includegraphics[width=3.0in]{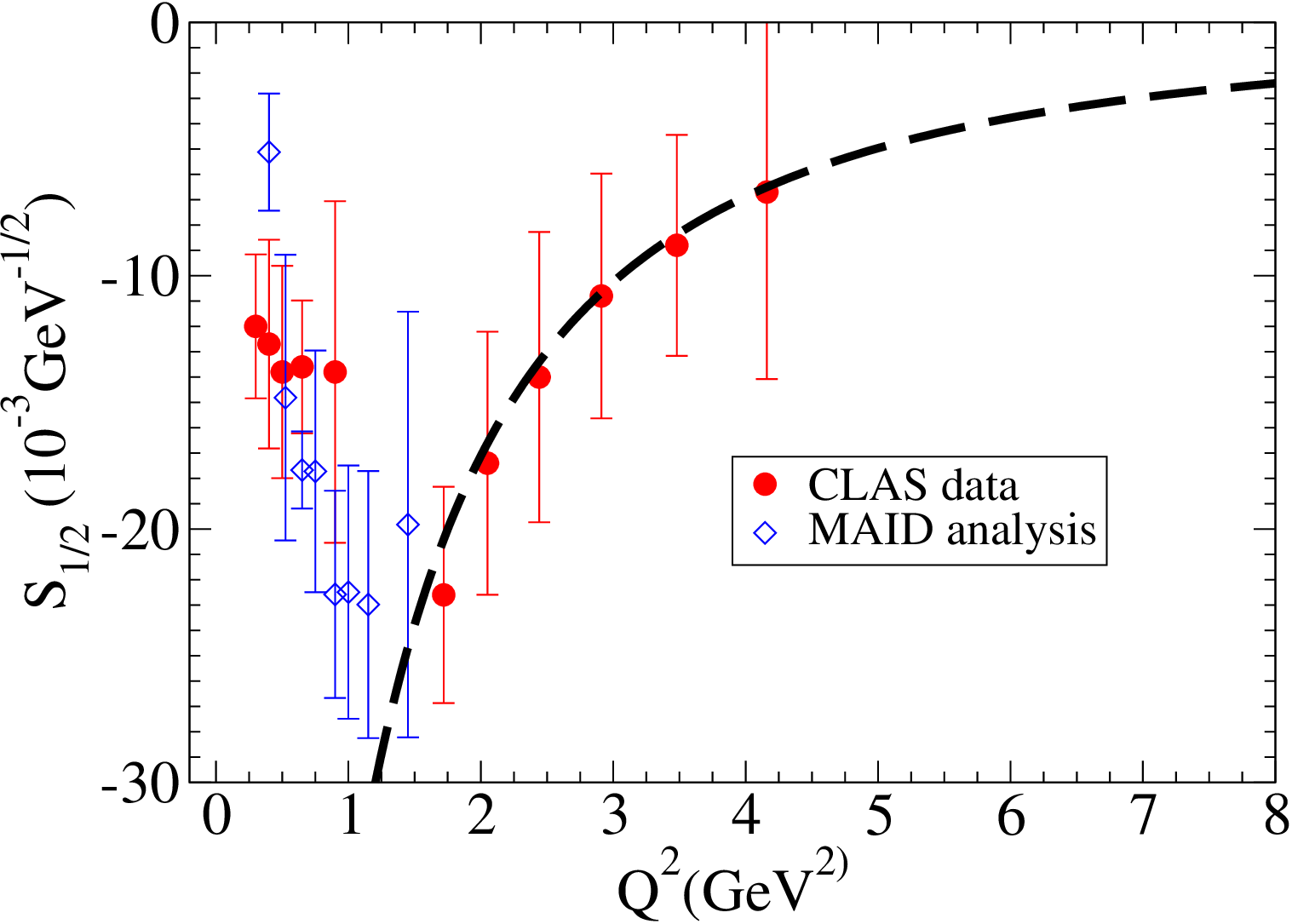}
  \caption{$\gamma N \to S_{11}(1525)$ helicity amplitudes~\cite{S11}
determined from the model for the form factors
with $F_2^\ast=0$. Data are from Refs.~\cite{CLAS,MAID,Dalton09}.}
\label{figAMP}
\end{figure}

%\newpage

\section{Results}

The results for the $\gamma N \to N^\ast(1535)$ transition 
form factors are~\cite{S11}:
\ba
F_1^\ast(Q^2)= \frac{1}{2}(3 j_1+ j_3) {\cal I}_0(Q^2), \hspace{1cm}
F_2^\ast(Q^2)= -\frac{1}{2}(3 j_2- j_4) \frac{M_S+M}{2M}
{\cal I}_0(Q^2),
\ea
where $j_i=\sfrac{1}{6}f_{i+}+ \sfrac{1}{2}f_{i-} \tau_3$
and  $j_{(i+2)}=\sfrac{1}{6}f_{i+}- \sfrac{1}{6}f_{i-} \tau_3$ ($i=1,2$),
and
${\cal I}_0(Q^2)$
the overlap integral between the scalar wave functions
that can be written in the $N^\ast(1535)$ rest frame as
\ba
{\cal I}_0(Q^2)= \int_k \frac{k_z}{|{\bf k}|} \psi_{S11}
(P_{S11},k) \psi_N(P_N,k),
\ea
where the factor $k_z$ is due to the $N^\ast(1535)$ P-state.
The expression for ${\cal I}_0$ can be used to define 
the applicable range of the model.
In the $Q^2 \to 0$ limit we can write ${\cal I}_0 (Q^2) \propto |{\bf q}|$,
where $|{\bf q}|$ is the photon 3-momentum in the
$N^\ast(1535)$ rest frame, which is
$|{\bf q}|= \sfrac{M_S^2-M^2}{2M_S}$, when $Q^2=0$.
As a consequence ${\cal I}_0(0) \ne 0$, if $M_S \ne M$,
meaning that nucleon and $N^\ast(1535)$ states are not orthogonal  \cite{S11}.
In a regime where  $|{\bf q}|=  \sfrac{M_S^2-M^2}{2M_S}$
is very small, one can regard that ${\cal I}_0(0) \approx 0$,
and the model is valid.
Taking $Q^2 \gg |{\bf q}|^2= 0.23$ GeV$^2$ one may assume that
the states are orthogonal in the region $Q^2 >$ 2.3 GeV$^2$.
In figures we will present our results for the region $Q^2>1$ GeV$^2$.

The results corresponding to the model
are presented in Fig.~\ref{figFF}, and compared
with the CLAS and MAID data~\cite{CLAS,MAID}, and
also with an estimate for the
valence quark core contributions from EBAC~\cite{EBAC}.
The results of the experimental data $F_2^\ast\simeq 0$,
for $Q^2 > 1.5$ GeV$^2$, are in contradiction with the
results of the model.
The simplest interpretation of this discrepancy
is that the valence quark effects
are canceled by the meson cloud contributions~\cite{S11}.
This interpretation is supported by EBAC~\cite{EBAC} and
effective chiral meson-baryon models~\cite{Jido08,Jido11}.
We use then the result $F_2^\ast=0$ in the calculation of
the helicity amplitudes $A_{1/2}$ and $S_{1/2}$.
The results are presented in Fig.~\ref{figAMP}.

\begin{figure}
% original size: [width=3.2in]
  \includegraphics[width=3.0in]{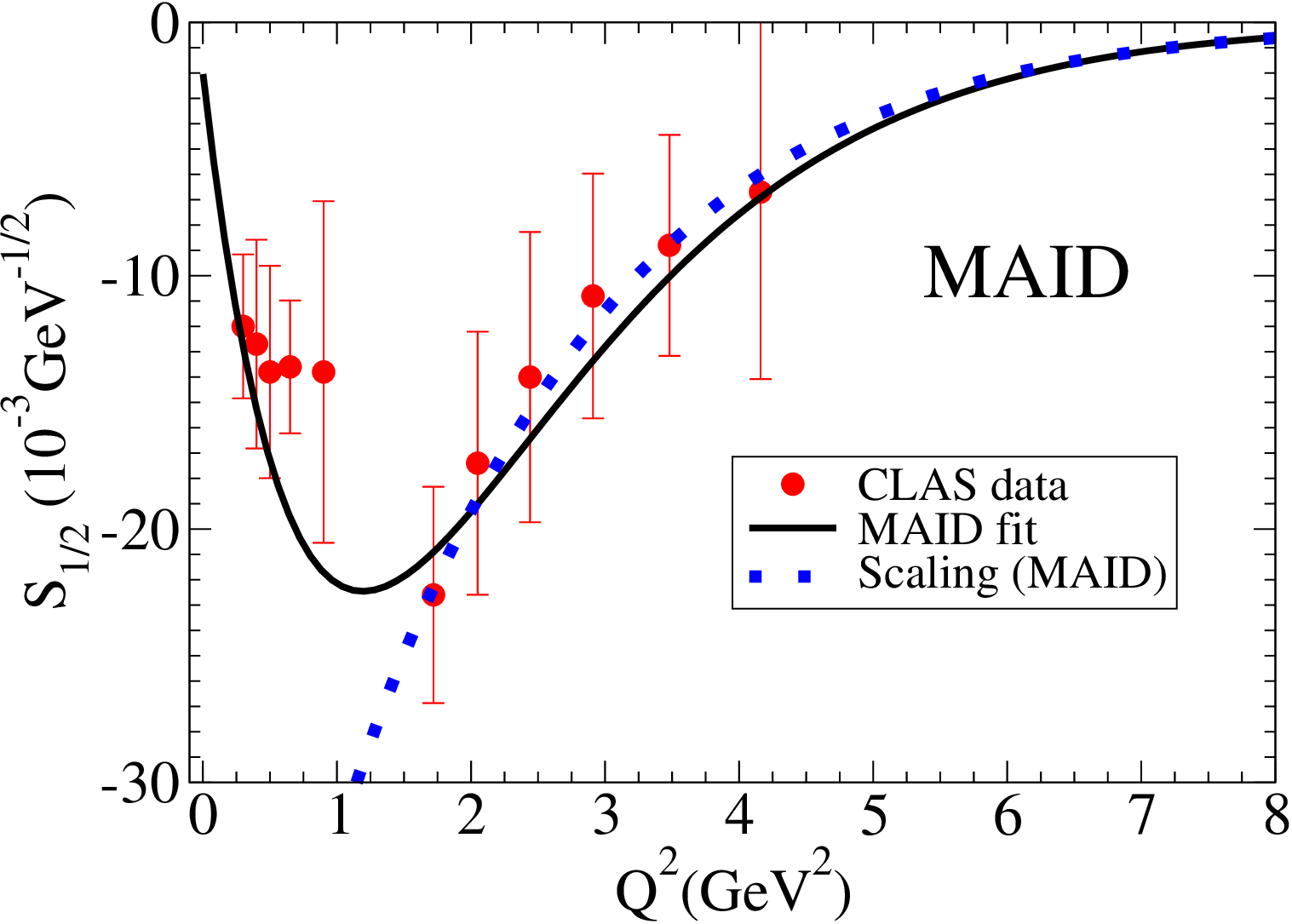}  \hspace{.1cm}
  \includegraphics[width=3.1in]{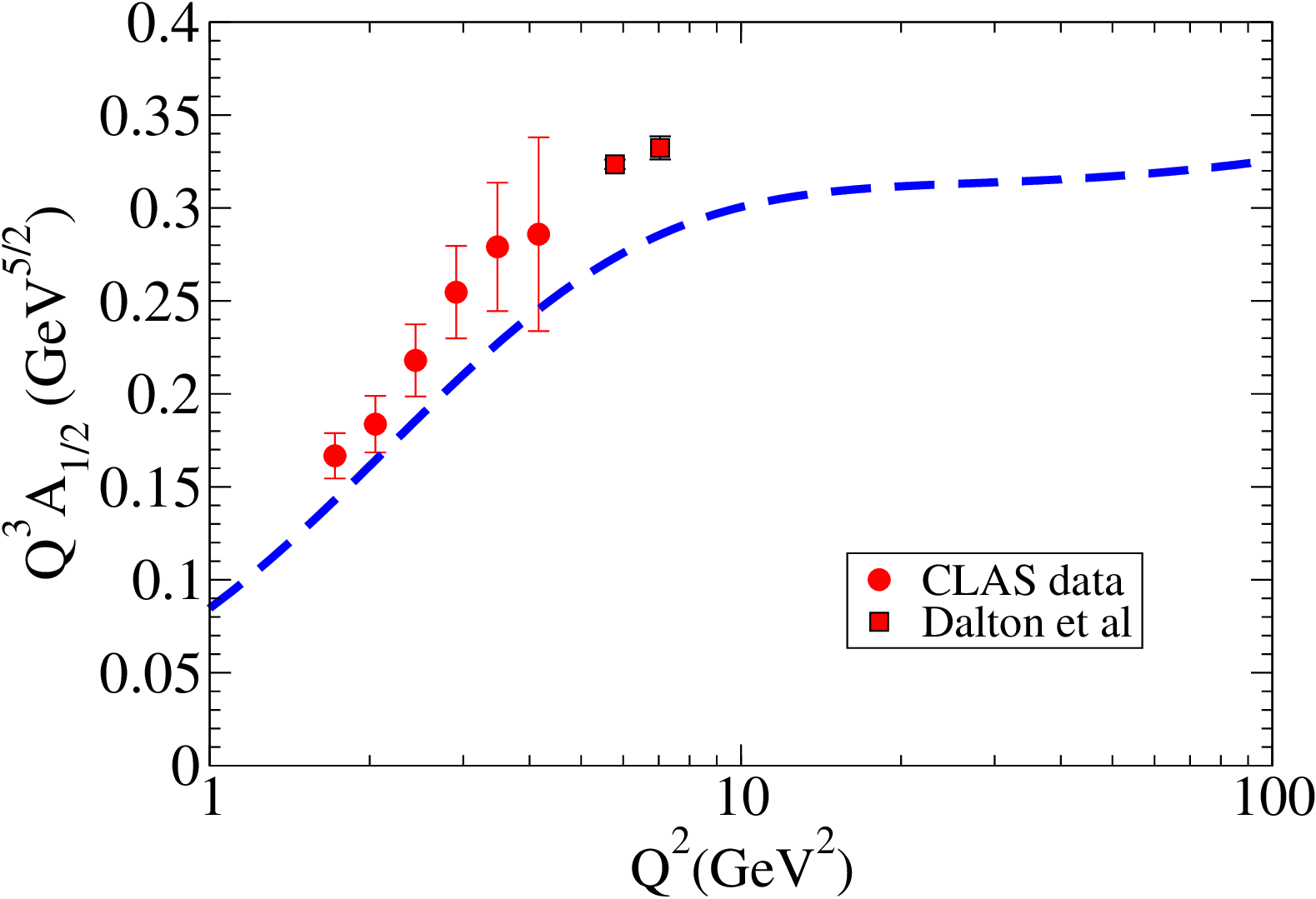}
  \caption{
Left panel: Test of the relation~(\ref{eqScaling}) using
the MAID parametrization for $A_{1/2}$ (doted line)
compared with the $S_{1/2}$ parametrization.
Right panel:
Comparison of the asymptotic behavior of the model for $A_{1/2}$
compared with the data~\cite{CLAS,Dalton09}.  }
\label{figMIX}
\end{figure}

The direct consequence of the result $F_2^\ast=0$
is the relation \cite{S11scaling}
\ba
S_{1/2}= - \frac{\sqrt{1+\tau}}{\sqrt{2}} \frac{M_S^2-M^2}{2 M_S Q}
A_{1/2},
\label{eqScaling}
\ea
where $\tau= \sfrac{Q^2}{(M_S+M)^2}$,
which holds for
$Q^2 > 1.8$ GeV$^2$ [when $|{\bf q}|\simeq Q \sqrt{1+ \tau}$].
We can see that the relation~(\ref{eqScaling})
is consistent with the data using our result for $A_{1/2}$,
and it is also valid for the MAID parametrization~\cite{MAID}
for $A_{1/2}$ and $S_{1/2}$ amplitudes~\cite{S11scaling}.
The test for MAID is presented in the left panel of Fig.~\ref{figMIX}.
Finally, we look the asymptotic behavior in $Q^2$
for $A_{1/2}$ and compare with the data.
Our result and the data are consistent with the $1/Q^3$ behavior, apart
from a small logarithmic correction (see the right panel of Fig.~\ref{figMIX}).

In conclusion the $\gamma N \to N^\ast(1535)$ reaction is a very
interesting reaction from the constituent quark model perspective.
The quark degrees of freedom are sufficient to
explain the $F_1^\ast$ data, 
but insufficient to explain the $F_2^\ast$ data for large $Q^2$.
Extracted data $F_2^\ast=0$ for large $Q^2$, leads to
a very interesting relation between the $S_{1/2}$ and $A_{1/2}$
helicity amplitudes given by Eq.~(\ref{eqScaling}).
When interpreted in terms of the valence quark
and the meson cloud excitation effects, Eq.~(\ref{eqScaling}) 
is the consequence of the cancellation between the two effects.
Accurate data for $A_{1/2}$ and $S_{1/2}$ for $Q^2> 2$ GeV$^2$,  
are necessary to test the relation~(\ref{eqScaling})
and clarify this point.
Efforts from quark models, dynamical coupled-channel models,
chiral effective models, QCD sum rules~\cite{Braun09}, 
and lattice QCD, are welcome in order to
interpret the $\gamma N \to N^\ast(1535)$ reaction data.

%%%%%%%%%%%%%%%%%%%%%%%%%%%%%%%%%%%%%%%%%%%%%%%%
%% BACKMATTER
%%%%%%%%%%%%%%%%%%%%%%%%%%%%%%%%%%%%%%%%%%%%%%%%

\begin{theacknowledgments}
G.~R.~was supported by the Funda\c{c}\~ao para
a Ci\^encia e a Tecnologia under the Grant
No.~SFRH/BPD/26886/2006.
K.~T.~was supported by the University of Adelaide and
the Australian Research Council through the Grant FL0992247 (AWT).
\end{theacknowledgments}



\bibliographystyle{aipproc}   % if natbib is available

%\bibliography{sample}

%\input{biblo}

\end{document}